\documentclass[aps,prl,reprint,groupedaddress]{revtex4-1}
\usepackage{graphicx}
\usepackage{amsmath}
\usepackage{mathrsfs}
\usepackage{array}
\usepackage[breaklinks=true,colorlinks=true,linkcolor=blue,urlcolor=blue,citecolor=blue]{hyperref}
\graphicspath{{./figures/}}

\begin{document}

\title{Determining Metastable Ion Lifetime and History through Wave-Particle Interaction} 
\author{F. Chu}
\email[]{feng-chu@uiowa.edu}
\affiliation{Department of Physics and Astronomy, University of Iowa, Iowa City, Iowa 52242, USA}
\author{F. Skiff}
\affiliation{Department of Physics and Astronomy, University of Iowa, Iowa City, Iowa 52242, USA}
\date{\today}

\begin{abstract}
Laser-induced fluorescence (LIF) performed on metastable ions is frequently used to probe the dynamics of ground-state ion motions in many laboratory plasmas. However, these measurements place restrictions on the metastable ion lifetime. Metastable states are produced from direct ionization of neutral atoms as well as ions in other electronic states, of which the former will only faithfully represent processes that act on the ion dynamics in a time shorter than the metastable lifetime. We present here the first experimental study of this type of systematic effect using wave-particle interaction in an Argon multidipole plasma. The metastable lifetime and relative fraction of metastables produced from pre-existing ions, necessary for correcting the LIF measurement errors, can be determined by fitting the experimental results with the theory we propose.
\end{abstract}

\maketitle

\textit{Introduction}.---Laser-induced fluorescence (LIF) is a nonintrusive, nominally nonperturbative diagnostic technique that has found application in the study of a wide range of fundamental and applied problems in plasma physics. For example, LIF is often adopted to probe the plasma parameters in the plume of Hall effect thrusters to study the physical processes that control their operation \cite{w._a._hargus_laser-induced_2001, mazouffre_time-resolved_2009}. In semiconductor fabrication, LIF is employed to measure the plasma ion velocity distribution in the silicon etching process \cite{jacobs_phase-resolved_2010}. A reliable phase-space diagnostic is also required in the study of plasma sheath formation \cite{severn_experimental_2003}, ion heating \cite{mcchesney_observation_1987}, and velocity-space diffusion \cite{bowles_direct_1992}.

In gas discharges, LIF is performed on metastable ions that are produced directly from neutral gas particles and also from ions in other electronic states \cite{cherrington_gaseous_1979, goeckner_laserinduced_1991}. Here rises an important question: when can Doppler-resolved LIF on metastable ions be used to infer the velocity distribution of ground-state ions (the majority ion population) in many laboratory plasmas? In principle, LIF measurements of any observable quantities derived from the ion velocity distribution are affected by this fundamental issue.


Previous experimental results \cite{nakano_metastable_1992, sadeghi_ion_1991} suggest that there are limitations of this laser diagnostic technique due to the finite lifetime of metastable ions. Simulations based on our newly developed Lagrangian model for LIF \cite{claire_nonlinear_2001, chu_determining_2017, chu_lagrangian_2018} show that under circumstances where the metastable ion population is produced from direct ionization of neutrals, the velocity distribution measured using LIF will only faithfully represent processes which act on the ion dynamics in a time shorter than the metastable lifetime. For instance in wave measurements \cite{sarfaty_direct_1996, mcwilliams_experimental_1986}, the perturbed distribution $f_1(v,t)$ on these metastables cannot be correct if the wave period is greater than the metastable lifetime. However, the LIF performed on the metastable population produced from pre-existing ions is not affected by the metastable lifetime. 


Understanding the behavior of each metastable population is crucial in LIF applications as it provides a guideline for avoiding the systematic errors caused by the finite metastable lifetime. In the case where these errors are inevitable, correction of the LIF measurements requires knowledge on the metastable lifetime and fraction of metastables produced from pre-existing ions as opposed to directly from neutral atoms. However, unlike other well-known systematic errors existing in LIF measurements such as optical pumping broadening \cite{goeckner_laserinduced_1989, goeckner_saturation_1993}, the metastable lifetime effects have never been explored experimentally before. In addition, it is a long-standing problem to trace the production history of metastable ions.

In this Letter, we report the first experimental measurement of metastable ion lifetime in a plasma as well as the relative fraction of metastables produced from pre-existing ions. The technique relies on measuring the ionic wave response. A theory is also presented to demonstrate that the LIF measurement errors can be corrected when the metastable lifetime effects become critical.



\textit{Theory and simulation}.---Laboratory plasmas are often in the regime where the ion sound speed is much larger than the ion thermal speed. If the neutrals and ions are assumed to have the same temperature (both neutrals and ions have the same zeroth-order distribution $f_0$), by solving the Vlasov equation perturbatively for weak electric field one can obtain the first- and second-order perturbation of the LIF measured ion distribution in the presence of an electrostatic wave
\begin{gather}
\label{eq:f1LIF}
f_{\textup{1-LIF}}=- \frac{iEe}{m_{\textup{i}}} \cdot \frac{\partial f_0}{\partial v}\left ( \frac{n_{\textup{meta-i}}}{\omega}+ \frac{n_{\textup{meta-n}}}{\omega+i\xi } \right ), \\[2ex]
\label{eq:f2LIF}
f_{\textup{2-LIF}}=- \left (  \frac{Ee}{m_{\textup{i}}} \right )^2 \cdot \frac{\partial^2 f_0}{\partial v^2}\left [ \frac{n_{\textup{meta-i}}}{\omega^2}+ \frac{n_{\textup{meta-n}}}{(\omega+i\xi)^2 } \right ],
\end{gather}
where $E$ is the amplitude of the electrostatic wave, $\omega$ is the wave angular frequency, $m_{\textup{i}}$ is the ion mass, and the densities of the metastable ions produced from neutrals and pre-existing ions are denoted by $n_{\textup{meta-n}}$ and $n_{\textup{meta-i}}$, respectively. The metastable lifetime $\tau$ is controlled by metastable quench rate $r$, electron-collisional excitation rate $u$, and optical pumping rate $W$
\begin{equation}
\label{eq:lifetime}
1/\tau=\xi=r+u+W.
\end{equation}
Though it is difficult to find the exact solutions of $f_{\textup{1-LIF}}$ and $f_{\textup{2-LIF}}$ in a general case where the neutral atoms and ions are not in thermal equilibrium, their numerical solutions can still be calculated using our Lagrangian model for LIF described bellow.


In the limit when the metastable lifetime is long compared to the wave period ($\tau \gg 1/\omega$), the LIF measured first-order perturbation $f_{\textup{1-LIF}}$ in Eq. (\ref{eq:f1LIF}) is proportional to the total metastable density $n_{\textup{meta-i}}+n_{\textup{meta-n}}$. On the contrary, when the metastable lifetime is short ($\tau \ll 1/\omega$), $f_{\textup{1-LIF}}$ is proportional to $n_{\textup{meta-i}}$ with no contribution from $n_{\textup{meta-n}}$ at all. The difference between these two metastable populations results from their distinctive histories. The lifetime of the metastables produced from neutrals sets the time scale they experience the wave field. With a history of being neutral particles, this population cannot react to the electric field until they become ions. If the lifetime is shorter than the wave period, these metastables will not live long enough to interact with the wave, resulting in a reduction in the measured $f_1$. On the other hand, as metastables produced from pre-existing ions have already fully interacted with the wave field before becoming metastables, the perturbed distribution measured using LIF is independent of their lifetime. This analysis can also be applied to $f_{\textup{2-LIF}}$ in Eq.~(\ref{eq:f2LIF}). 


\begin{figure}[b]
\begin{center}
\includegraphics[width=3.3in]{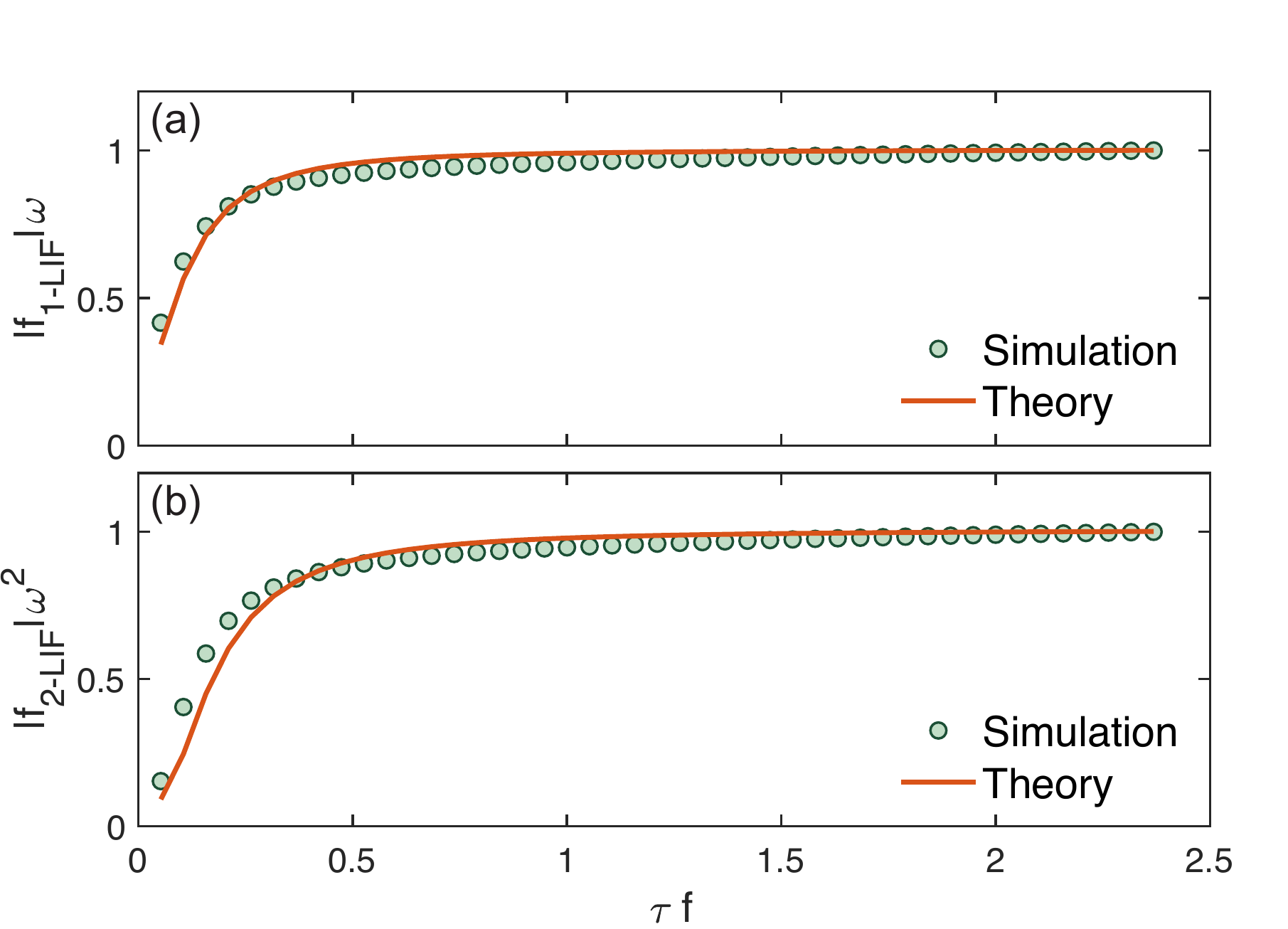}
\caption{Comparison between the Lagrangian model and the theory for $f_{\textup{1-LIF}}$ and $f_{\textup{2-LIF}}$ at various metastable lifetimes is shown in (a) and (b), respectively. The metastable lifetime $\tau$ is normalized by the wave frequency $f=\omega/2\pi$. The relative fraction of metastables produced from pre-existing ions $n_{\textup{meta-i}}/(n_{\textup{meta-i}}+n_{\textup{meta-n}})=14.3$ \%.}
\label{fig:simulation-theory}
\end{center}
\end{figure}

A numerical simulation based on a Lagrangian model for LIF is performed to test the theory for $f_{\textup{1-LIF}}$ and $f_{\textup{2-LIF}}$. In the Lagrangian interpretation, one must follow each individual ion orbit as it moves through space and time. This approach achieves a large computational advantage by exploiting the separation of the classical dynamics of the ions from the quantum mechanics of the electronic states, reducing a system of coupled partial differential equations in the traditional Eulerian model to ordinary differential equations. Furthermore, since this model does not impose constraints on the ion orbits, it can be applied to systems with complicated ion dynamics. A detailed description of the model and its application are presented in \cite{chu_determining_2017} and \cite{chu_lagrangian_2018}. The simulation results for $f_{\textup{1-LIF}}$ and $f_{\textup{2-LIF}}$ demonstrate a good agreement with the theoretical predictions, as shown in Fig. \ref{fig:simulation-theory}.

The amplitude of the electrostatic wave can be computed from Eqs. (\ref{eq:f1LIF})--(\ref{eq:f2LIF}) as
\begin{equation}
\label{eq:ELIF}
E_{\textup{LIF}}= \left | \left (f_{\textup{2-LIF}}\cdot \frac{\partial f_0}{\partial v} \right ) \bigg/ \left ( f_{\textup{1-LIF}}\cdot \frac{\partial^2 f_0}{\partial v^2} \right ) \right | \cdot \frac{m_{\textup{i}}\omega }{e}.
\end{equation}
As expected, the LIF measured wave amplitude $E_{\textup{LIF}}$ is subject to the metastable lifetime effects. If the same electric field can be measured using a different method which does not rely on metastable ions, such as an electric field probe, then the metastable lifetime effects can be observed experimentally by comparing the results from these two measurements.


\textit{Experiment}.---We demonstrate the technique to measure the metastable lifetime and history in an Argon plasma confined in a multidipole chamber of 73 cm length and 49 cm diameter \cite{hood_ion_2016}. A hot cathode consisting of lanthanum-hexaboride (LaB6) heated by a resistive graphite bar is biased at $-70$ V with respect to the chamber walls, emitting primary electrons to produce the plasma through impact ionization. Emission current from the cathode is regulated at 56 mA. The multidipole confinement is provided by an electrically grounded magnet cage consisting of 16 rows of magnets with alternating poles covering all inside walls of the chamber. The magnetic field strength is about 1000 G on the surface of the magnets and less than 2 G in the bulk plasma.


\begin{figure*}
\begin{center}
\includegraphics[width=6.4in]{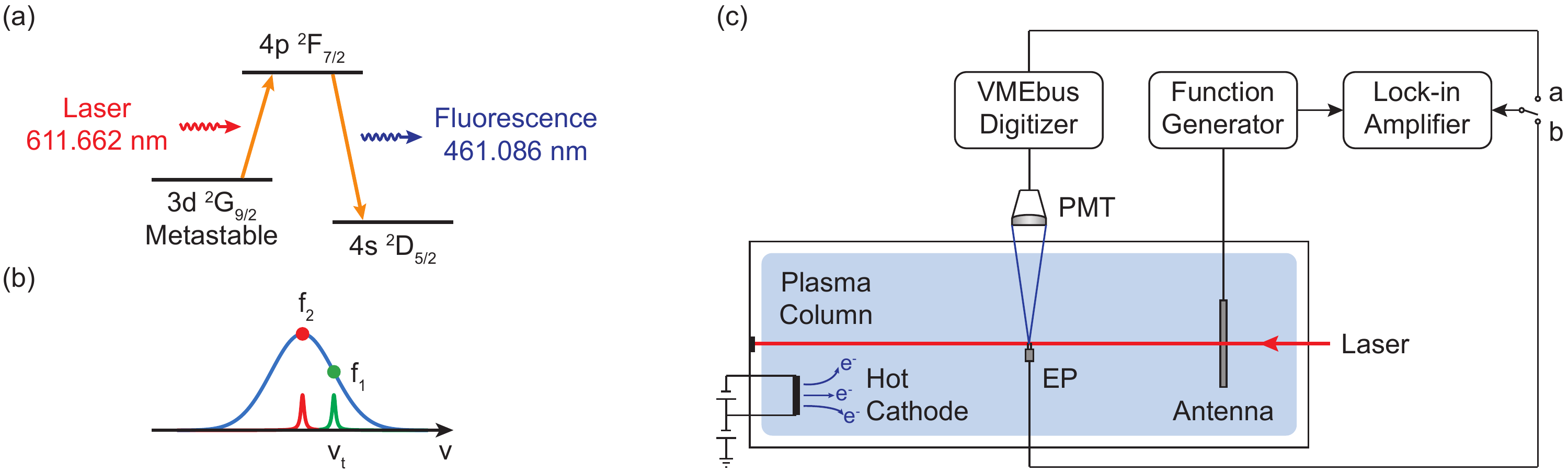}
\caption{(a) Energy level diagram of the laser-induced fluorescence process. (b) Frequencies of the laser selected to resolve $f_1$ and $f_2$ in velocity-space in the experiment; $f_1$ is measured at thermal velocity $v_\textup{t}$ and $f_2$ at the peak of $f_0$. (c) Schematic of the experimental setup. Neutral pressure is regulated by a mass flow controller and measured using an ionization gauge. The LIF diagnostic equipment (not drawn to scale) includes a double-mesh antenna with a diameter of 3.81 cm and 65 $\%$ open area, a 16-channel photomultiplier tube (PMT), a Versa Module Europa bus (VMEbus) board, and a lock-in amplifier. A disc-shaped Langmuir probe with a diameter of 0.65 cm is placed in the bulk plasma to measure the electron density and temperature. A differential sinusoidal signal with $V_\textup{p}=\pm2$ V is applied on the antenna that is 5 cm away from the LIF viewing volume to excite ion acoustic waves in the plasma. The wave electric field is measured using both LIF and a double-tip electric field probe (EP).}
\label{fig:exp}
\end{center}
\end{figure*}



Figure~\ref{fig:exp}(a) shows the LIF scheme used in the experiment. It is accomplished by a single mode tunable Rhodamine 6G dye laser (Sirah Matisse-DS). To induce fluorescence, in the rest frame of an ion, the laser is tuned at 611.662 nm to excite electrons in the $\textup{3d }^2\textup{G}_{9/2}$ metastable state to the $\textup{4p }^2\textup{F}_{7/2}^{o}$ state. Fluorescence photons are emitted at 461.086 nm when those electrons decay to the $\textup{4s }^2\textup{D}_{5/2}$ state with a large branching ratio of $66.5$ \% \cite{severn_argon_1998, mattingly_measurement_2013}. In principle the LIF measured electric field $E_{\textup{LIF}}$ in Eq. (\ref{eq:ELIF}) can be obtained by sampling $f_1$ and $f_2$ at almost any point in velocity-space. However, to achieve a better signal-to-noise ratio the LIF measurements of $f_1$ and $f_2$ are made at $v_\textup{t}$ (ion thermal speed) and the peak of $f_0$ respectively, as illustrated in Fig.~\ref{fig:exp}(b).


The experimental setup is depicted in Fig.~\ref{fig:exp}(c). An ion acoustic wave is generated in the plasma by applying a differential sinusoidal signal with $V_\textup{p}=\pm2$ V on the double-mesh antenna. This driving signal is sent to the lock-in amplifier as well as a reference. The frequency of the wave is scanned from 1 kHz to 45 kHz with an increment of 1 kHz. At each wave frequency $\omega$, $f_1$ and $f_2$ are resolved using the lock-in amplifier by locking the frequency at $\omega$ and $2\omega$ respectively. The LIF measured electric field $E_{\textup{LIF}}$ can therefore be calculated using Eq. (\ref{eq:ELIF}). The same electric field is also evaluated using a double-tip electric field probe to compare with the LIF measurements.


The electric field probe is made with a low noise, high speed instrumentation amplifier AD8421 which allows to extract low level differential voltage signals in the presence of high frequency common-mode noise over a wide frequency range \cite{berumen_analysis_2018}. Since plasmas tend to have a large impedance, even small capacitance from the wires connecting the probe tips and the instrumentation amplifier can significantly reduce the bandwidth of the probe. Therefore, the instrumentation amplifier is placed only 15 mm away from the tips to improve the probe's performance in the high frequency range. The electric field probe measures the differential voltage between two points in the direction of the wave propagation and gives the electric field $E_{\textup{probe}}=V_\textup{out}/dG$, where $V_\textup{out}$ is the output voltage of the probe and $G=100$ is the gain of the instrumentation amplifier. The separation $d$ between the probe tips is about half a millimeter, providing an excellent spatial resolution in the electric field measurements. The measured electric field $E_{\textup{probe}}$ also needs to be corrected by multiplying a factor $\alpha$ to compensate the errors mainly caused by the large plasma impedance comparable to the input impedance of the instrumentation amplifier and difficulties in precisely measuring the tip separation $d$. 


\textit{Results}.---By scanning the laser wavelength, it is found that the ions have a Maxwellian velocity distribution along the direction of the laser beam in the center of the chamber. The ion temperature $T_\textup{i}=0.03 \pm 0.01$ eV, suggesting that both ions and neutrals are close to the room temperature 0.025 eV. The ion thermal speed is given by $v_\textup{t} = \sqrt{T_\textup{i}/m_\textup{i}} \approx 2.70 \times 10^4$ $\textup{cm/s}$. The other plasma parameters are measured using a disc-shaped Langmuir probe. At neutral pressure $P=0.058 \pm 0.006$ mTorr, the typical parameters in the bulk are, electron density $n_\textup{e}=2.10 \times 10^9$ $\textup{cm}^{-3}$, electron temperature $T_\textup{e}=2.90$ eV, and plasma potential $V_\textup{p}=-4.30$ V. The ion sound speed is estimated as $C_\textup{s} \approx \sqrt{T_\textup{e}/m_\textup{i}}=2.60 \times 10^5$ $\textup{cm/s}$, which is much larger than the ion thermal speed $v_\textup{t}$. Therefore both assumptions made in the derivation of Eqs. (\ref{eq:f1LIF})--(\ref{eq:f2LIF}) are satisfied.

\begin{figure*}
\begin{center}
\includegraphics[width=6.6in]{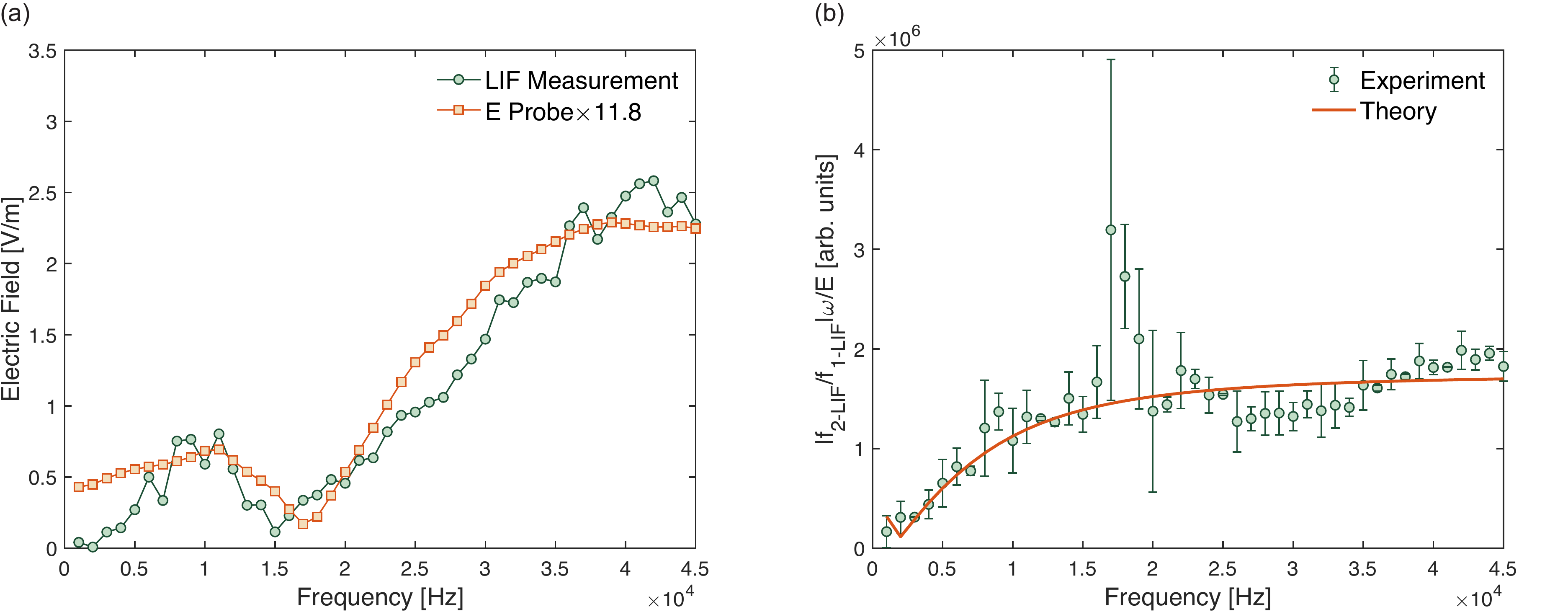}
\caption{(a) Comparison of the ion acoustic wave electric field measured using LIF and the electric field probe at different frequencies. The probe measurement $E_{\textup{probe}}$ is multiplied by a correction factor $\alpha=11.8$ to scale with the LIF measurement $E_{\textup{LIF}}$. (b) LIF measurement of the wave electric field $E_{\textup{LIF}}$ normalized by the probe measurement $E_{\textup{probe}}$. Error bars represent one-standard-deviation uncertainties. The theoretical prediction is also plotted here for comparison.}
\label{fig:f2-f1-E}
\end{center}
\end{figure*}

The comparison between $E_{\textup{LIF}}$ and $E_{\textup{probe}}$, the electric field of the ion acoustic wave measured using LIF and the electric field probe respectively, is presented in Fig.~\ref{fig:f2-f1-E}(a). The probe measurement is multiplied by a correction factor $\alpha=11.8$ to scale with the LIF measurement. For a small electric field, the probe's electronic pickup of fast electrons accelerated by the antenna can introduce significant errors to the measurement, causing the dips on the two curves to slightly shift from each other around 16 kHz. The electric field measured using the two methods are in good agreement above 10 kHz, however, the LIF measurement is systematically smaller than the probe measurement below 10 kHz due to the metastable lifetime effects.


To be able to compare the experiment with the theory in Eq. (\ref{eq:ELIF}), the LIF measured electric field $E_{\textup{LIF}}$ needs to be normalized by the probe measurement $E_{\textup{probe}}$. The result of this procedure, shown in Fig.~\ref{fig:f2-f1-E}(b), is the key experimental result of this Letter. The peak at 16 kHz results from the misalignment of the dips in Fig.~\ref{fig:f2-f1-E}(a). The low-frequency roll-off evident in Fig.~\ref{fig:f2-f1-E}(b) is due to the finite metastable lifetime. Both the metastable lifetime and fraction of metastables produced from pre-existing ions can be determined by fitting the experimental data with the theory. Because of the coulomb collisional drag effect (not present for $v$ at the peak of $f_0$), ions with $v \sim v_\textup{t}$ systematically spend less time in resonance with the laser, reducing the optical pumping ($W$ in our computational model). From the best fit, it is found that the inverse metastable lifetime for ions at the peak of $f_0$ is $\xi=(5.63 \pm 0.35) \times 10^4$ $\textup{s}^{-1}$, which gives the metastable lifetime $\tau=17.8 \pm 1.1$ $\mu \textup{s}$. Similarly, at ion thermal velocity $\xi=(4.09 \pm 0.23) \times 10^4$ $\textup{s}^{-1}$ and $\tau=24.4 \pm 1.4$ $\mu \textup{s}$. The sum of the quench rate and collisional excitation rate $r+u$ is estimated as $1 \times 10^4$ $\textup{s}^{-1}$ which is at least three times smaller than the optical pumping rate $W$, making the latter the dominant factor in controlling the metastable lifetime in the experiment. The relative fraction of the metastables produced from pre-existing ions $n_{\textup{meta-i}}/(n_{\textup{meta-i}}+n_{\textup{meta-n}})=4 \pm 2$ \%, suggesting that the metastable ions are mainly produced by direct ionization of neutrals in this Argon multidipole plasma \cite{goeckner_laserinduced_1991}.

The theoretical quench rate and collisional excitation rate are also computed to compare with the experimental values. The quench rate is given by $r=n_{\textup{n}} \sigma \sqrt{8T_{\textup{n}}/\pi m_{\textup{n}}}=1.9 \times 10^3$ $\textup{s}^{-1}$, where $n_{\textup{n}}$ is the neutral density, $\sigma=2.7 \times 10^{-14}$ $\textup{cm}^{2}$ \cite{skiff_ion_2001} is the quench cross section, $T_{\textup{n}}$ is the neutral temperature in energy units, and $m_{\textup{n}}$ is the mass of the neutral particles. The collisional excitation rate is estimated as $u \approx 1 \times 10^4$ $\textup{s}^{-1}$ \cite{curry_measurement_1995}. The sum of the two rates can then be calculated as $1.2 \times 10^4$ $\textup{s}^{-1}$, which is consistent with our experimental value within errors.

\textit{Conclusions}.---We have presented the first experimental study of the metastable lifetime effects using wave-particle interaction and LIF in a multidipole plasma. The experimental finding verifies that LIF performed on metastable ions produced directly from neutral atoms can only be used to infer the velocity distribution of ground-state ions if the ion dynamics is in a time shorter than the metastable lifetime. By fitting the theory with the experiment results, the metastable lifetime and relative fraction of metastables produced from pre-existing ions can be determined. Under circumstances where the metastable lifetime effects are inevitable, in our case probing the wave electric field under 10 kHz with LIF, the measurement errors can be corrected using the theory addressed in this Letter. Lastly, we demonstrate that LIF measurements of $f_1$ and $f_2$ provide a new method to determine the absolute electric field in a plasma. Since this technique does not perturb the local field, it can be used to calibrate other electric field measurement tools, such as the double-tip electric field probe used in our experiment.




We acknowledge R. Hood for designing and constructing the electric field probe and the double-mesh antenna. This work was supported by the U.S. Department of Energy under Grant No. DE-SC0016473.

\bibliography{refs}

\begin{thebibliography}{23}%
\makeatletter
\providecommand \@ifxundefined [1]{%
 \@ifx{#1\undefined}
}%
\providecommand \@ifnum [1]{%
 \ifnum #1\expandafter \@firstoftwo
 \else \expandafter \@secondoftwo
 \fi
}%
\providecommand \@ifx [1]{%
 \ifx #1\expandafter \@firstoftwo
 \else \expandafter \@secondoftwo
 \fi
}%
\providecommand \natexlab [1]{#1}%
\providecommand \enquote  [1]{``#1''}%
\providecommand \bibnamefont  [1]{#1}%
\providecommand \bibfnamefont [1]{#1}%
\providecommand \citenamefont [1]{#1}%
\providecommand \href@noop [0]{\@secondoftwo}%
\providecommand \href [0]{\begingroup \@sanitize@url \@href}%
\providecommand \@href[1]{\@@startlink{#1}\@@href}%
\providecommand \@@href[1]{\endgroup#1\@@endlink}%
\providecommand \@sanitize@url [0]{\catcode `\\12\catcode `\$12\catcode
  `\&12\catcode `\#12\catcode `\^12\catcode `\_12\catcode `\%12\relax}%
\providecommand \@@startlink[1]{}%
\providecommand \@@endlink[0]{}%
\providecommand \url  [0]{\begingroup\@sanitize@url \@url }%
\providecommand \@url [1]{\endgroup\@href {#1}{\urlprefix }}%
\providecommand \urlprefix  [0]{URL }%
\providecommand \Eprint [0]{\href }%
\providecommand \doibase [0]{http://dx.doi.org/}%
\providecommand \selectlanguage [0]{\@gobble}%
\providecommand \bibinfo  [0]{\@secondoftwo}%
\providecommand \bibfield  [0]{\@secondoftwo}%
\providecommand \translation [1]{[#1]}%
\providecommand \BibitemOpen [0]{}%
\providecommand \bibitemStop [0]{}%
\providecommand \bibitemNoStop [0]{.\EOS\space}%
\providecommand \EOS [0]{\spacefactor3000\relax}%
\providecommand \BibitemShut  [1]{\csname bibitem#1\endcsname}%
\let\auto@bib@innerbib\@empty
\bibitem [{\citenamefont {W.~A.~Hargus}\ and\ \citenamefont
  {Cappelli}(2001)}]{w._a._hargus_laser-induced_2001}%
  \BibitemOpen
  \bibfield  {author} {\bibinfo {author} {\bibfnamefont {J.}~\bibnamefont
  {W.~A.~Hargus}}\ and\ \bibinfo {author} {\bibfnamefont {M.~A.}\ \bibnamefont
  {Cappelli}},\ }\href {\doibase 10.1007/s003400100589} {\bibfield  {journal}
  {\bibinfo  {journal} {Appl Phys B}\ }\textbf {\bibinfo {volume} {72}},\
  \bibinfo {pages} {961} (\bibinfo {year} {2001})}\BibitemShut {NoStop}%
\bibitem [{\citenamefont {Mazouffre}\ \emph {et~al.}(2009)\citenamefont
  {Mazouffre}, \citenamefont {Gawron},\ and\ \citenamefont
  {Sadeghi}}]{mazouffre_time-resolved_2009}%
  \BibitemOpen
  \bibfield  {author} {\bibinfo {author} {\bibfnamefont {S.}~\bibnamefont
  {Mazouffre}}, \bibinfo {author} {\bibfnamefont {D.}~\bibnamefont {Gawron}}, \
  and\ \bibinfo {author} {\bibfnamefont {N.}~\bibnamefont {Sadeghi}},\ }\href
  {\doibase 10.1063/1.3112704} {\bibfield  {journal} {\bibinfo  {journal}
  {Physics of Plasmas}\ }\textbf {\bibinfo {volume} {16}},\ \bibinfo {pages}
  {043504} (\bibinfo {year} {2009})}\BibitemShut {NoStop}%
\bibitem [{\citenamefont {Jacobs}\ \emph {et~al.}(2010)\citenamefont {Jacobs},
  \citenamefont {Gekelman}, \citenamefont {Pribyl},\ and\ \citenamefont
  {Barnes}}]{jacobs_phase-resolved_2010}%
  \BibitemOpen
  \bibfield  {author} {\bibinfo {author} {\bibfnamefont {B.}~\bibnamefont
  {Jacobs}}, \bibinfo {author} {\bibfnamefont {W.}~\bibnamefont {Gekelman}},
  \bibinfo {author} {\bibfnamefont {P.}~\bibnamefont {Pribyl}}, \ and\ \bibinfo
  {author} {\bibfnamefont {M.}~\bibnamefont {Barnes}},\ }\href {\doibase
  10.1103/PhysRevLett.105.075001} {\bibfield  {journal} {\bibinfo  {journal}
  {Phys. Rev. Lett.}\ }\textbf {\bibinfo {volume} {105}},\ \bibinfo {pages}
  {075001} (\bibinfo {year} {2010})}\BibitemShut {NoStop}%
\bibitem [{\citenamefont {Severn}\ \emph {et~al.}(2003)\citenamefont {Severn},
  \citenamefont {Wang}, \citenamefont {Ko},\ and\ \citenamefont
  {Hershkowitz}}]{severn_experimental_2003}%
  \BibitemOpen
  \bibfield  {author} {\bibinfo {author} {\bibfnamefont {G.~D.}\ \bibnamefont
  {Severn}}, \bibinfo {author} {\bibfnamefont {X.}~\bibnamefont {Wang}},
  \bibinfo {author} {\bibfnamefont {E.}~\bibnamefont {Ko}}, \ and\ \bibinfo
  {author} {\bibfnamefont {N.}~\bibnamefont {Hershkowitz}},\ }\href {\doibase
  10.1103/PhysRevLett.90.145001} {\bibfield  {journal} {\bibinfo  {journal}
  {Phys. Rev. Lett.}\ }\textbf {\bibinfo {volume} {90}},\ \bibinfo {pages}
  {145001} (\bibinfo {year} {2003})}\BibitemShut {NoStop}%
\bibitem [{\citenamefont {McChesney}\ \emph {et~al.}(1987)\citenamefont
  {McChesney}, \citenamefont {Stern},\ and\ \citenamefont
  {Bellan}}]{mcchesney_observation_1987}%
  \BibitemOpen
  \bibfield  {author} {\bibinfo {author} {\bibfnamefont {J.~M.}\ \bibnamefont
  {McChesney}}, \bibinfo {author} {\bibfnamefont {R.~A.}\ \bibnamefont
  {Stern}}, \ and\ \bibinfo {author} {\bibfnamefont {P.~M.}\ \bibnamefont
  {Bellan}},\ }\href {\doibase 10.1103/PhysRevLett.59.1436} {\bibfield
  {journal} {\bibinfo  {journal} {Phys. Rev. Lett.}\ }\textbf {\bibinfo
  {volume} {59}},\ \bibinfo {pages} {1436} (\bibinfo {year}
  {1987})}\BibitemShut {NoStop}%
\bibitem [{\citenamefont {Bowles}\ \emph {et~al.}(1992)\citenamefont {Bowles},
  \citenamefont {McWilliams},\ and\ \citenamefont {Rynn}}]{bowles_direct_1992}%
  \BibitemOpen
  \bibfield  {author} {\bibinfo {author} {\bibfnamefont {J.}~\bibnamefont
  {Bowles}}, \bibinfo {author} {\bibfnamefont {R.}~\bibnamefont {McWilliams}},
  \ and\ \bibinfo {author} {\bibfnamefont {N.}~\bibnamefont {Rynn}},\ }\href
  {\doibase 10.1103/PhysRevLett.68.1144} {\bibfield  {journal} {\bibinfo
  {journal} {Phys. Rev. Lett.}\ }\textbf {\bibinfo {volume} {68}},\ \bibinfo
  {pages} {1144} (\bibinfo {year} {1992})}\BibitemShut {NoStop}%
\bibitem [{\citenamefont {Cherrington}(1979)}]{cherrington_gaseous_1979}%
  \BibitemOpen
  \bibfield  {author} {\bibinfo {author} {\bibfnamefont {B.~E.}\ \bibnamefont
  {Cherrington}},\ }\href@noop {} {\emph {\bibinfo {title} {Gaseous Electronics
  and Gas Lasers}}}\ (\bibinfo  {publisher} {{Pergamon Press}},\ \bibinfo
  {year} {1979})\BibitemShut {NoStop}%
\bibitem [{\citenamefont {Goeckner}\ \emph {et~al.}(1991)\citenamefont
  {Goeckner}, \citenamefont {Goree},\ and\ \citenamefont
  {Sheridan}}]{goeckner_laserinduced_1991}%
  \BibitemOpen
  \bibfield  {author} {\bibinfo {author} {\bibfnamefont {M.~J.}\ \bibnamefont
  {Goeckner}}, \bibinfo {author} {\bibfnamefont {J.}~\bibnamefont {Goree}}, \
  and\ \bibinfo {author} {\bibfnamefont {T.~E.}\ \bibnamefont {Sheridan}},\
  }\href {\doibase 10.1063/1.859924} {\bibfield  {journal} {\bibinfo  {journal}
  {Physics of Fluids B: Plasma Physics (1989-1993)}\ }\textbf {\bibinfo
  {volume} {3}},\ \bibinfo {pages} {2913} (\bibinfo {year} {1991})}\BibitemShut
  {NoStop}%
\bibitem [{\citenamefont {Nakano}\ \emph {et~al.}(1992)\citenamefont {Nakano},
  \citenamefont {Sadeghi}, \citenamefont {Trevor}, \citenamefont {Gottscho},\
  and\ \citenamefont {Boswell}}]{nakano_metastable_1992}%
  \BibitemOpen
  \bibfield  {author} {\bibinfo {author} {\bibfnamefont {T.}~\bibnamefont
  {Nakano}}, \bibinfo {author} {\bibfnamefont {N.}~\bibnamefont {Sadeghi}},
  \bibinfo {author} {\bibfnamefont {D.~J.}\ \bibnamefont {Trevor}}, \bibinfo
  {author} {\bibfnamefont {R.~A.}\ \bibnamefont {Gottscho}}, \ and\ \bibinfo
  {author} {\bibfnamefont {R.~W.}\ \bibnamefont {Boswell}},\ }\href {\doibase
  10.1063/1.351460} {\bibfield  {journal} {\bibinfo  {journal} {Journal of
  Applied Physics}\ }\textbf {\bibinfo {volume} {72}},\ \bibinfo {pages} {3384}
  (\bibinfo {year} {1992})}\BibitemShut {NoStop}%
\bibitem [{\citenamefont {Sadeghi}\ \emph {et~al.}(1991)\citenamefont
  {Sadeghi}, \citenamefont {Nakano}, \citenamefont {Trevor},\ and\
  \citenamefont {Gottscho}}]{sadeghi_ion_1991}%
  \BibitemOpen
  \bibfield  {author} {\bibinfo {author} {\bibfnamefont {N.}~\bibnamefont
  {Sadeghi}}, \bibinfo {author} {\bibfnamefont {T.}~\bibnamefont {Nakano}},
  \bibinfo {author} {\bibfnamefont {D.~J.}\ \bibnamefont {Trevor}}, \ and\
  \bibinfo {author} {\bibfnamefont {R.~A.}\ \bibnamefont {Gottscho}},\ }\href
  {\doibase 10.1063/1.350332} {\bibfield  {journal} {\bibinfo  {journal}
  {Journal of Applied Physics}\ }\textbf {\bibinfo {volume} {70}},\ \bibinfo
  {pages} {2552} (\bibinfo {year} {1991})}\BibitemShut {NoStop}%
\bibitem [{\citenamefont {Claire}\ \emph {et~al.}(2001)\citenamefont {Claire},
  \citenamefont {Dindelegan}, \citenamefont {Bachet},\ and\ \citenamefont
  {Skiff}}]{claire_nonlinear_2001}%
  \BibitemOpen
  \bibfield  {author} {\bibinfo {author} {\bibfnamefont {N.}~\bibnamefont
  {Claire}}, \bibinfo {author} {\bibfnamefont {M.}~\bibnamefont {Dindelegan}},
  \bibinfo {author} {\bibfnamefont {G.}~\bibnamefont {Bachet}}, \ and\ \bibinfo
  {author} {\bibfnamefont {F.}~\bibnamefont {Skiff}},\ }\href {\doibase
  10.1063/1.1419221} {\bibfield  {journal} {\bibinfo  {journal} {Review of
  Scientific Instruments}\ }\textbf {\bibinfo {volume} {72}},\ \bibinfo {pages}
  {4372} (\bibinfo {year} {2001})}\BibitemShut {NoStop}%
\bibitem [{\citenamefont {Chu}\ \emph {et~al.}(2017)\citenamefont {Chu},
  \citenamefont {Mattingly}, \citenamefont {Berumen}, \citenamefont {Hood},\
  and\ \citenamefont {Skiff}}]{chu_determining_2017}%
  \BibitemOpen
  \bibfield  {author} {\bibinfo {author} {\bibfnamefont {F.}~\bibnamefont
  {Chu}}, \bibinfo {author} {\bibfnamefont {S.~W.}\ \bibnamefont {Mattingly}},
  \bibinfo {author} {\bibfnamefont {J.}~\bibnamefont {Berumen}}, \bibinfo
  {author} {\bibfnamefont {R.}~\bibnamefont {Hood}}, \ and\ \bibinfo {author}
  {\bibfnamefont {F.}~\bibnamefont {Skiff}},\ }\href {\doibase
  10.1088/1748-0221/12/11/C11005} {\bibfield  {journal} {\bibinfo  {journal}
  {J. Inst.}\ }\textbf {\bibinfo {volume} {12}},\ \bibinfo {pages} {C11005}
  (\bibinfo {year} {2017})}\BibitemShut {NoStop}%
\bibitem [{\citenamefont {Chu}\ and\ \citenamefont
  {Skiff}(2018)}]{chu_lagrangian_2018}%
  \BibitemOpen
  \bibfield  {author} {\bibinfo {author} {\bibfnamefont {F.}~\bibnamefont
  {Chu}}\ and\ \bibinfo {author} {\bibfnamefont {F.}~\bibnamefont {Skiff}},\
  }\href {\doibase 10.1063/1.5020088} {\bibfield  {journal} {\bibinfo
  {journal} {Physics of Plasmas}\ }\textbf {\bibinfo {volume} {25}},\ \bibinfo
  {pages} {013506} (\bibinfo {year} {2018})}\BibitemShut {NoStop}%
\bibitem [{\citenamefont {Sarfaty}\ \emph {et~al.}(1996)\citenamefont
  {Sarfaty}, \citenamefont {Souza-Machado},\ and\ \citenamefont
  {Skiff}}]{sarfaty_direct_1996}%
  \BibitemOpen
  \bibfield  {author} {\bibinfo {author} {\bibfnamefont {M.}~\bibnamefont
  {Sarfaty}}, \bibinfo {author} {\bibfnamefont {S.~D.}\ \bibnamefont
  {Souza-Machado}}, \ and\ \bibinfo {author} {\bibfnamefont {F.}~\bibnamefont
  {Skiff}},\ }\href {\doibase 10.1063/1.871581} {\bibfield  {journal} {\bibinfo
   {journal} {Physics of Plasmas (1994-present)}\ }\textbf {\bibinfo {volume}
  {3}},\ \bibinfo {pages} {4316} (\bibinfo {year} {1996})}\BibitemShut
  {NoStop}%
\bibitem [{\citenamefont {McWilliams}\ and\ \citenamefont
  {Sheehan}(1986)}]{mcwilliams_experimental_1986}%
  \BibitemOpen
  \bibfield  {author} {\bibinfo {author} {\bibfnamefont {R.}~\bibnamefont
  {McWilliams}}\ and\ \bibinfo {author} {\bibfnamefont {D.}~\bibnamefont
  {Sheehan}},\ }\href {\doibase 10.1103/PhysRevLett.56.2485} {\bibfield
  {journal} {\bibinfo  {journal} {Phys. Rev. Lett.}\ }\textbf {\bibinfo
  {volume} {56}},\ \bibinfo {pages} {2485} (\bibinfo {year}
  {1986})}\BibitemShut {NoStop}%
\bibitem [{\citenamefont {Goeckner}\ and\ \citenamefont
  {Goree}(1989)}]{goeckner_laserinduced_1989}%
  \BibitemOpen
  \bibfield  {author} {\bibinfo {author} {\bibfnamefont {M.~J.}\ \bibnamefont
  {Goeckner}}\ and\ \bibinfo {author} {\bibfnamefont {J.}~\bibnamefont
  {Goree}},\ }\href {\doibase 10.1116/1.575831} {\bibfield  {journal} {\bibinfo
   {journal} {Journal of Vacuum Science \& Technology A}\ }\textbf {\bibinfo
  {volume} {7}},\ \bibinfo {pages} {977} (\bibinfo {year} {1989})}\BibitemShut
  {NoStop}%
\bibitem [{\citenamefont {Goeckner}\ \emph {et~al.}(1993)\citenamefont
  {Goeckner}, \citenamefont {Goree},\ and\ \citenamefont
  {Sheridan}}]{goeckner_saturation_1993}%
  \BibitemOpen
  \bibfield  {author} {\bibinfo {author} {\bibfnamefont {M.~J.}\ \bibnamefont
  {Goeckner}}, \bibinfo {author} {\bibfnamefont {J.}~\bibnamefont {Goree}}, \
  and\ \bibinfo {author} {\bibfnamefont {T.~E.}\ \bibnamefont {Sheridan}},\
  }\href {\doibase 10.1063/1.1144103} {\bibfield  {journal} {\bibinfo
  {journal} {Review of Scientific Instruments}\ }\textbf {\bibinfo {volume}
  {64}},\ \bibinfo {pages} {996} (\bibinfo {year} {1993})}\BibitemShut
  {NoStop}%
\bibitem [{\citenamefont {Hood}\ \emph {et~al.}(2016)\citenamefont {Hood},
  \citenamefont {Scheiner}, \citenamefont {Baalrud}, \citenamefont {Hopkins},
  \citenamefont {Barnat}, \citenamefont {Yee}, \citenamefont {Merlino},\ and\
  \citenamefont {Skiff}}]{hood_ion_2016}%
  \BibitemOpen
  \bibfield  {author} {\bibinfo {author} {\bibfnamefont {R.}~\bibnamefont
  {Hood}}, \bibinfo {author} {\bibfnamefont {B.}~\bibnamefont {Scheiner}},
  \bibinfo {author} {\bibfnamefont {S.~D.}\ \bibnamefont {Baalrud}}, \bibinfo
  {author} {\bibfnamefont {M.~M.}\ \bibnamefont {Hopkins}}, \bibinfo {author}
  {\bibfnamefont {E.~V.}\ \bibnamefont {Barnat}}, \bibinfo {author}
  {\bibfnamefont {B.~T.}\ \bibnamefont {Yee}}, \bibinfo {author} {\bibfnamefont
  {R.~L.}\ \bibnamefont {Merlino}}, \ and\ \bibinfo {author} {\bibfnamefont
  {F.}~\bibnamefont {Skiff}},\ }\href {\doibase 10.1063/1.4967870} {\bibfield
  {journal} {\bibinfo  {journal} {Physics of Plasmas}\ }\textbf {\bibinfo
  {volume} {23}},\ \bibinfo {pages} {113503} (\bibinfo {year}
  {2016})}\BibitemShut {NoStop}%
\bibitem [{\citenamefont {Severn}\ \emph {et~al.}(1998)\citenamefont {Severn},
  \citenamefont {Edrich},\ and\ \citenamefont
  {McWilliams}}]{severn_argon_1998}%
  \BibitemOpen
  \bibfield  {author} {\bibinfo {author} {\bibfnamefont {G.~D.}\ \bibnamefont
  {Severn}}, \bibinfo {author} {\bibfnamefont {D.~A.}\ \bibnamefont {Edrich}},
  \ and\ \bibinfo {author} {\bibfnamefont {R.}~\bibnamefont {McWilliams}},\
  }\href {\doibase 10.1063/1.1148472} {\bibfield  {journal} {\bibinfo
  {journal} {Review of Scientific Instruments}\ }\textbf {\bibinfo {volume}
  {69}},\ \bibinfo {pages} {10} (\bibinfo {year} {1998})}\BibitemShut {NoStop}%
\bibitem [{\citenamefont {Mattingly}\ \emph {et~al.}(2013)\citenamefont
  {Mattingly}, \citenamefont {Berumen}, \citenamefont {Chu}, \citenamefont
  {Hood},\ and\ \citenamefont {Skiff}}]{mattingly_measurement_2013}%
  \BibitemOpen
  \bibfield  {author} {\bibinfo {author} {\bibfnamefont {S.~W.}\ \bibnamefont
  {Mattingly}}, \bibinfo {author} {\bibfnamefont {J.}~\bibnamefont {Berumen}},
  \bibinfo {author} {\bibfnamefont {F.}~\bibnamefont {Chu}}, \bibinfo {author}
  {\bibfnamefont {R.}~\bibnamefont {Hood}}, \ and\ \bibinfo {author}
  {\bibfnamefont {F.}~\bibnamefont {Skiff}},\ }\href {\doibase
  10.1088/1748-0221/8/11/C11015} {\bibfield  {journal} {\bibinfo  {journal} {J.
  Inst.}\ }\textbf {\bibinfo {volume} {8}},\ \bibinfo {pages} {C11015}
  (\bibinfo {year} {2013})}\BibitemShut {NoStop}%
\bibitem [{\citenamefont {Berumen}\ and\ \citenamefont
  {Skiff}(2018)}]{berumen_analysis_2018}%
  \BibitemOpen
  \bibfield  {author} {\bibinfo {author} {\bibfnamefont {J.}~\bibnamefont
  {Berumen}}\ and\ \bibinfo {author} {\bibfnamefont {F.}~\bibnamefont
  {Skiff}},\ }\href {\doibase 10.1063/1.5058805} {\bibfield  {journal}
  {\bibinfo  {journal} {Physics of Plasmas}\ }\textbf {\bibinfo {volume}
  {25}},\ \bibinfo {pages} {122102} (\bibinfo {year} {2018})}\BibitemShut
  {NoStop}%
\bibitem [{\citenamefont {Skiff}\ \emph {et~al.}(2001)\citenamefont {Skiff},
  \citenamefont {Bachet},\ and\ \citenamefont {Doveil}}]{skiff_ion_2001}%
  \BibitemOpen
  \bibfield  {author} {\bibinfo {author} {\bibfnamefont {F.}~\bibnamefont
  {Skiff}}, \bibinfo {author} {\bibfnamefont {G.}~\bibnamefont {Bachet}}, \
  and\ \bibinfo {author} {\bibfnamefont {F.}~\bibnamefont {Doveil}},\ }\href
  {\doibase 10.1063/1.1379044} {\bibfield  {journal} {\bibinfo  {journal}
  {Physics of Plasmas}\ }\textbf {\bibinfo {volume} {8}},\ \bibinfo {pages}
  {3139} (\bibinfo {year} {2001})}\BibitemShut {NoStop}%
\bibitem [{\citenamefont {Curry}\ \emph {et~al.}(1995)\citenamefont {Curry},
  \citenamefont {Skiff}, \citenamefont {Sarfaty},\ and\ \citenamefont
  {Good}}]{curry_measurement_1995}%
  \BibitemOpen
  \bibfield  {author} {\bibinfo {author} {\bibfnamefont {J.~J.}\ \bibnamefont
  {Curry}}, \bibinfo {author} {\bibfnamefont {F.}~\bibnamefont {Skiff}},
  \bibinfo {author} {\bibfnamefont {M.}~\bibnamefont {Sarfaty}}, \ and\
  \bibinfo {author} {\bibfnamefont {T.~N.}\ \bibnamefont {Good}},\ }\href
  {\doibase 10.1103/PhysRevLett.74.1767} {\bibfield  {journal} {\bibinfo
  {journal} {Phys. Rev. Lett.}\ }\textbf {\bibinfo {volume} {74}},\ \bibinfo
  {pages} {1767} (\bibinfo {year} {1995})}\BibitemShut {NoStop}%
\end{thebibliography}%

\end{document}